\documentstyle[aps,prl,psfig,twocolumn]{revtex}
\newcommand{\avg}[1]{\langle{#1}\rangle}

\begin{document}
\narrowtext
{\bf Comment on: Thermal model for Adaptive Competition in a Market:}
Cavagna {\em et al.} \cite{CGGS} introduced an interesting model -- called
TMG as in [1] -- 
similar to the minority game \cite{MG,Savit} (MG), of $N$ agents 
interacting in a market. Strategies of agents are represented by $D$ 
dimensional vectors $\vec{R}_i^a$ with $i=1,\ldots, N$ running through
agents and $a=1,\ldots, s$ through $i^{\rm th}$ agent's available
choices. The strategy $\vec{R}^\star_i$ used by $i$ is selected 
drawing
$a=\star$ from a Boltzmann distribution given by Eq. (4) of ref. [1] --
or (4)-[1] for short -- with ``temperature'' $T$
and energies $-P(\vec{R}^a_i)$ (see Eq. (3)-[1]).
Cavagna {\em et al.} [1] report numerical data showing an interesting 
collective behavior 
as a function of $T$ (figs. 2-[1] and 3-[1]) and arrive at 
Eqs. (5,6)-[1] which are claimed to be the ``exact dynamical 
equations for'' the TMG. We show here 
that {\em i)} Eqs. (5,6)-[1] are incorrect {\em ii)} the correct 
continuum time dynamics is the same as that of the MG \cite{CMZ}. 
As a consequence the
analytic solution of the MG of ref. \cite{CMZ} holds also
for the TMG. Finally 
{\em iii)} the features found in [1] for $T\gg 1$ (figs. 
2,3-[1]) are due to small simulation times and disappears
if the system is in a steady state.

Cavagna {\em et al.} fail to define properly the continuum time limit (CTL) 
prescription, which is essential for stochastic differential equations
such as Eq. (5)-[1]. It is crucial, in a proper derivation of the CTL, to 
observe that characteristic times in the TMG are proportional to $D$, as shown 
numerically in Fig. 1. 
This is natural because the adaptation of each agent's strategy requires an 
optimization of all its $D$ components. This need sampling $\sim D$ 
values of $\vec\eta$, i.e. a time of order $D$. In order to eliminate
the dependence of times on system size $N=D/d$, one has to rescale time
by a factor $D$. 
The dynamics in the rescaled time $\tau=t/D$ is obtained iterating 
Eq. (3)-[1] from $t=D\tau$ to $D\tau'$
\begin{equation}\FL
\frac{P(\vec{R},\tau')-P(\vec{R},\tau)}{\tau'-\tau}
=\frac{-d}{D(\tau'\!-\!\tau)}\sum_{t=D\tau}^{D\tau'-1}
A(t)\,\vec R\cdot\vec\eta(t).
\label{eq1}
\end{equation}
The law of large numbers implies that, when $D=dN\to\infty$, 
the r.h.s. converges to $d\avg{A\,\vec R\cdot\vec\eta}$ where the average
$\avg{\ldots}$ is both on the distribution $\pi_i^a$ of 
$\vec{R}^\star_i$ {\em and} on that of $\vec{\eta}$. 
If we then let $\tau'\to\tau$
the l.h.s. converges to the derivative $\dot P$ of $P$ w.r.t. $\tau$.
Hence, using Eq. (2)-[1] 
for $A(t)$ and $\avg{\eta_{\alpha}\eta_\beta}=\delta_{\alpha,\beta}/D$, 
Eq. (\ref{eq1}) becomes 
$\dot P=-\frac{1}{N}\sum_i\avg{{\vec R}_i^\star}\cdot\vec R$ with
$\avg{{\vec R}_i^\star}=\sum_a\pi_i^a\vec R_i^a$. 
The combination of Eq. (\ref{eq1}) and Eq. (4)-[1] yields 
a dynamic equation for $\pi_i^a$, which reads
\begin{equation}\FL
\dot\pi_i^a=-\frac{1}{NT}\pi_i^a\sum_{j=1}^N\avg{{\vec R}_j^\star}
\cdot\left({\vec R}_i^a-\avg{{\vec R}_i^\star}\right).
\label{eq2}
\end{equation}
Eq. (2) coincides with the continuum time
equation of ref. \cite{CMZ} which leads to the 
exact solution of the MG for $N\to\infty$. This depends
only on the first two moments of the distribution of the components
of $\vec R_i^a$, which plays the role of quenched disorder.
Since, in the TMG, $\avg{\avg{\vec R_i^a}}=0$ and 
$\avg{\avg{(\vec R^a_i)^2}}=D$, these are the same 
as in the MG. Hence the two models have exactly 
the same collective behavior, as confirmed by Fig. 1.

Eq. (\ref{eq2}) suggests that the dependence on $T$ disappears by
time rescaling. This is true in the $d\ge d_c$ phase: 
The $T$ dependence for $T\gg 1$ reported in Figs 2,3-[1] is an 
artifact due to short simulation times (see also ref. \cite{BDD}).
For $d<d_c$ the CTL only holds for $T$ larger than a cross-over
$T_c(d)$, as discussed elsewhere \cite{CM}. Indeed for $d=0.1<d_c$
and $T$ large enough, data nicely collapses onto a single curve (see 
inset) once plotted against $\tau/T$. For $T<T_c(d)$ the solution 
of Eq. (\ref{eq2}) becomes dynamically unstable and the
system enters into a turbulent regime where the CTL breaks down \cite{CM}.

\begin{figure}
\centerline{\psfig{file=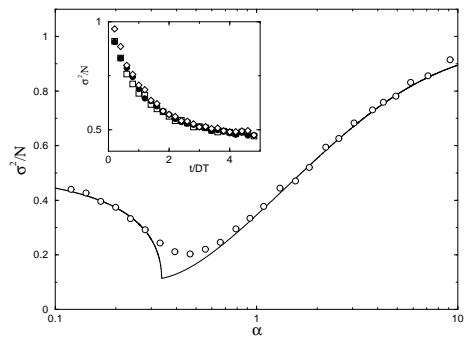,width=6cm,angle=270}}
\caption{$\sigma^2/N$ as a function of $d$:
numerical data with $S=2$, $D=64$ and $T=10$ ($\circ$)
and analytic solution [4] (full line). Finite size effects
occur close to the phase transition $d\approx d_c$. Inset:
relaxation of $\sigma^2/N$ for $d=0.1$, $DT=10^5$ and $D=25$ 
($\bullet$) $D=50$ ($\Box$) and $D=100$ ($\diamond$).
Data collapse implies that characteristic times are 
proportional to $D$ and $T$.}
\label{fig1}
\end{figure}

\noindent
D. Challet$^{(1)}$,
M. Marsili$^{(2)}$, 
and R. Zecchina$^{(3)}$\\
(1) Institut de Physique Th\'eorique, 
Universit\'e de Fribourg, 
P\'erolles, Fribourg, CH-1700, Switzerland.\\
(2) Istituto Nazionale per la Fisica della Materia,
SISSA unit, V. Beirut 4, I-34014 Trieste.\\
(3) Abdus Salam International Center for Theoretical Physics,
Strada Costiera, 11, I-34100, Trieste.

\end{document}